# Quantifying the intrinsic surface charge density and charge-transfer resistance of the graphene-solution interface through bias-free low-level charge measurement

*Jinglei Ping & A. T. Charlie Johnson[*]*

Department of Physics and Astronomy, University of Pennsylvania, Philadelphia, Pennsylvania 19104, United States

Liquid-based bio-applications of graphene require a quantitative understanding of the graphene-liquid interface, with the surface charge density of adsorbed ions, the interfacial charge transfer resistance, and the interfacial charge noise being of particular importance. We quantified these properties through measurements of the zero-bias Faradaic charge-transfer between graphene electrodes and aqueous solutions of varying ionic strength using a reproducible, low-noise, minimally perturbative charge measurement technique. The measurements indicated that adsorbed ions had a negative surface charge density of approximately -32.8 mC m$^{-2}$ and that the specific charge transfer resistance was 6.5 ± 0.3 MΩ cm$^2$. The normalized current noise power spectral density for all ionic concentrations tested collapsed onto a 1/f$^\alpha$ characteristic with α=1.1±0.2. All the results are in excellent agreement with predictions of the theory for the graphene-solution interface. This minimally-perturbative method for monitoring charge-transfer at the sub-pC scale exhibits low noise and ultra-low power consumption (~ fW), making it well-suited for use in low-level bioelectronics in liquid environments.

Over the past decades, investigations of low-dimensional nanomaterials for ultra-low-power bio-applications[1] in liquid environments have flourished[2, 3]. Graphene, a single atom-thick material with all atoms exposed to the environment, responds sensitively, rapidly, and with low noise[4] to variations in the local electrostatic/electrochemical potential. This makes graphene suitable for use in transduction elements (e.g., electrodes[5] and field-effect transistors[6]) to monitor charge accumulation and transfer processes in ionic solution[7]. Furthermore, large-area graphene[8] prepared by chemical vapor deposition (CVD)[9, 10] is amenable to top-down, scalable patterning using conventional techniques, enabling cost-effective fabrication of large arrays of miniaturized ultrasensitive bio-electrodes and biosensors[8, 11].

To enable liquid-based applications of graphene, it is important to quantify the basic physical properties of the graphene-solution interface, including the surface charge density of adsorbed ions, $\sigma_s$, and the interfacial charge-transfer resistance, $R_{ct}$. The surface charge density has profound effects on biochemical processes at the interface[12, 13], while the charge transfer resistance characterizes the kinetics at the interface and reflects the standard rate constant $k_0$[14]. Nonetheless, methods such as potentiometric titration[15] and surface plasmon resonance[13], have only recently been applied to quantify the surface charge density of graphene, and even the polarity of adsorbed ions remains a subject of debate[7, 16-19]. Conventional amperometric methods used to quantify $R_{ct}$, such as electrochemical impedance spectroscopy (EIS)[20], require perturbative ac voltage differences across the solid-liquid interface, so $R_{ct}$ is shunted by the impedance of the frequency-dependent constant phase element, and the inferred value of $R_{ct}$ is model-dependent[21]. Further, $R_{ct}$ is typically large for graphene (the specific contact resistance can be ~$10^6$ $\Omega$ cm$^2$)[22] and easily affected by surface contamination, surface roughness[21, 23, 24], and non-uniform current distribution[21, 25-27] induced by the applied ac voltage, making the value of $R_{ct}$ difficult to determine precisely[21].

In this report we quantify $\sigma_s$ and $R_{ct}$ through measurements of zero-bias, low-level (~ pC) Faradaic charge transfer at the graphene-solution interface using a non-perturbative electrometer-based approach[28]. We found that adsorbed ions had a negative surface charge density of approximately -32.8 mC per square meter and that the specific charge transfer resistance was 6.7 ± 0.7 MΩ cm$^2$. In addition, we observed characteristic 1/f noise in the Faradaic current in agreement with theoretical predictions for an ideal, unperturbed graphene-aqueous interface with slow Faradaic charge transfer[29, 30]. The approach has high accuracy and low noise, and it defines a pathway towards new classes of electronic nano-biodevices and nano-biosensors.

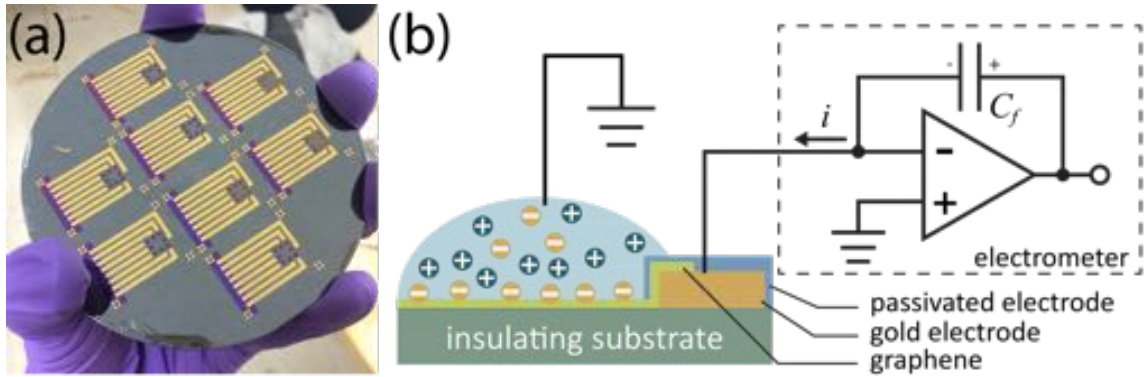

**Figure 1**. (a) Graphene-electrode devices fabricated on a 4-inch wafer. (b) Charge measurement configuration based on a feedback electrometer.

Fig. 1(a) is a photograph of graphene electrode devices used in these experiments. For the fabrication, large-area graphene is prepared via chemical vapor deposition[10, 31] and transferred using a low-contamination bubbling method onto a Si/SiO$_2$ substrate with pre-fabricated 45-nm thick Cr/Au electrodes. Next, 100 μm × 100 μm graphene electrodes are defined with photolithography and plasma etching. Metal contacts to the graphene electrodes are covered with a passivation layer consisting of ~ 20 μm thick SU-8 (MicroChem, Inc.) so that charge transfer occurs only at the graphene-solution interface. Further details of the fabrication process are provided in the Methods section.

The experimental setup is shown in Fig. 1(b). The noninverting input of the electrometer (Keithley 6517a) is initially grounded, and the inverting input is connected to a dissipating resistor. When a measurement commences (defined as time t=0), the inverting input is disconnected from the dissipating resistor and connected to the graphene electrode in phosphate buffer solution (pH=7). Since the potential at the inverting input evolves to match the potential at the noninverting input, the readout of the electrometer is the negative of the total charge induced in the graphene by the ions adsorbed on the graphene surface. As time increases, charge that is transferred from the solution to the graphene accumulates on the feedback capacitor $C_f$, and the electrometer readout changes linearly in time at a rate that varies systematically with the ionic strength of the solution (Fig. 2).

Care was taken to ensure that the noise level is well controlled in the measurement setup. Low-noise triaxial cables (Keithley 7078-TRX) were used to connect the electrometer with the device to reduce the current noise to the fA scale, according to the cable specification. Since the electrometer used a noiseless feedback capacitor, thermal noise associated with dissipative resistors is also avoided. For these reasons, the noise level of our setup approaches the minimum level set by the input bias current of the amplifier of the electrometer, 0.75 fA.

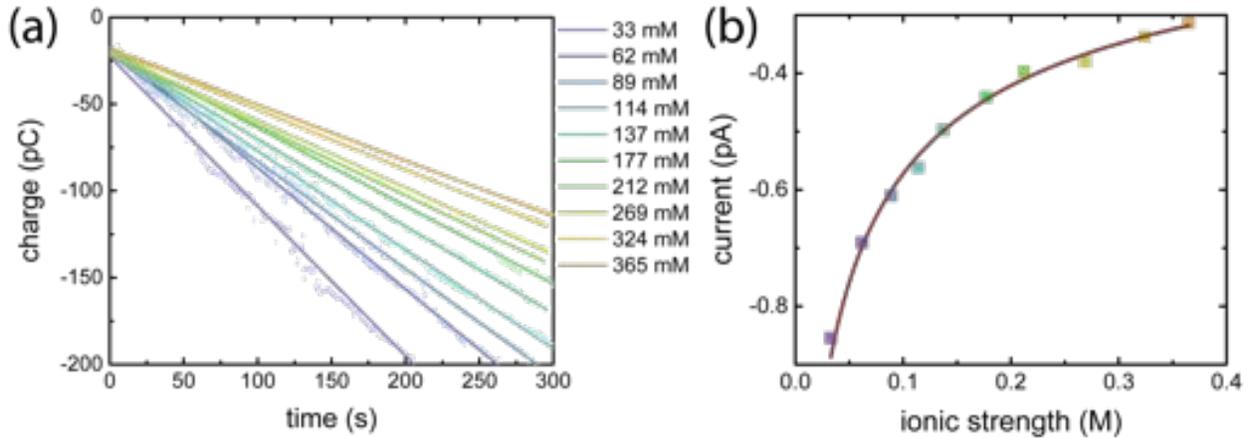

**Figure 2**. (a) Real-time measurement of charge transfer to a graphene electrode for solutions of ionic strength 33 - 365 mM. Solid lines are linear fits to the data. (b) Current flow to the graphene electrode from the solution, as a function of ionic strength. The solid curve is a fit to the data using the formula based on the theory for electric double layer. The sizes of the error bars in (a) and (b) are smaller than the sizes of the data points.

The real-time charge measurements in Fig. 2(a) demonstrate two characteristic features. First, a charge of approximately - 18.5 pC flows to the electrometer within 1 sec after the measurement starts, indicating that positive carriers (holes) with a charge density $\sigma_i$ of 1.8 mC m$^{-2}$ (number density of $1.1 \times 10^{12}$ cm$^{-2}$) accumulate in the grounded graphene electrode in response to contact with aqueous solution. This value is in good agreement with the change in carrier density inferred from the shift of the Dirac voltage of a similarly fabricated graphene field-effect transistor (GFET) exposed to ionic solution (~ 1.6 mC m$^{-2}$ or ~ $9.7 \times 10^{11}$ cm$^{-2}$) as shown in Fig. S1 of the Supplemental Material at [32]. From the polarity of the induced carriers (positive), we conclude that negative ions absorb on the graphene surface; the magnitude of the surface charge density will be determined below.

A second feature of the data in Fig. 2(a) is that the charge measured by the electrometer decreases linearly with time. This is attributed to Faradaic transfer of negative charge (electrons) from the ionic solution to the graphene through the charge-transfer resistance $R_{ct}$. The charge-transfer current *i* flowing through the graphene-solution interface is inferred from the slope of the charge-time data and plotted as a function of ionic strength in Fig. 2(b). The magnitude of the current, ~ 0.1 – 1.0 pA, is significantly smaller than the nA to μA current variation typically observed for back-gated GFET devices (see Fig. S1 of the Supplemental Material[26]), so the impact of charge transfer between the graphene and the ionic solution is negligible in liquid-based applications of back-gated GFET sensors[6]. However, in the liquid-gated GFET configuration, application of a gate voltage can cause the Faradaic current to be enhanced by orders of magnitude[22], to the point where it is comparable to the in-plane current[33].

The magnitude of the charge transfer current is observed to *decrease* with increasing ionic strength *c* (Fig. 2(b)). Incorporating the theory of the electric double layer (EDL)[34], the Faradaic charge transfer current is given by:

$$i = -\frac{1}{R_{ct}} \frac{2k_B T}{e} \sinh^{-1}\left(\frac{\sigma_d}{\sqrt{8\varepsilon\varepsilon_0 k_B T c}}\right) \qquad (1)$$

where $k_B$ is Boltzmann's constant, *T* is the absolute temperature, *e* is the electronic charge, $\varepsilon$ ($\varepsilon_0$) is the relative (vacuum) permittivity, and $\sigma_d$ is the areal charge density of the diffuse layer. (See S2 of the Supplemental Material[26] for discussion of the EDL) The *i-c* data is fit extremely well using Eq. (1), as shown in Fig. 3(a), with fit parameters $\sigma_d$ = 31 ± 3 mC m$^{-2}$ (~1.9 ± 0.2 × 10$^{13}$ cm$^{-2}$) and $R_{ct}$ = 67 ± 7 GΩ (equivalent to a specific resistance of 6.7 ± 0.7 MΩ cm$^2$). The best fit value of $\sigma_d$ is ~ 17 times larger than that of $\sigma_i$, indicating that the charge adsorbed on graphene is mostly neutralized by counter-ions in the solution, in good agreement with conclusions of fundamental electrochemical theory[16]. The total surface charge density of adsorbed ions on graphene, which is neutralized by the charge in the diffuse layer and charge carriers in graphene, is thus found to be $\sigma_s = -(\sigma_d + \sigma_i) = -32.8 \pm 3$ mC m$^{-2}$.

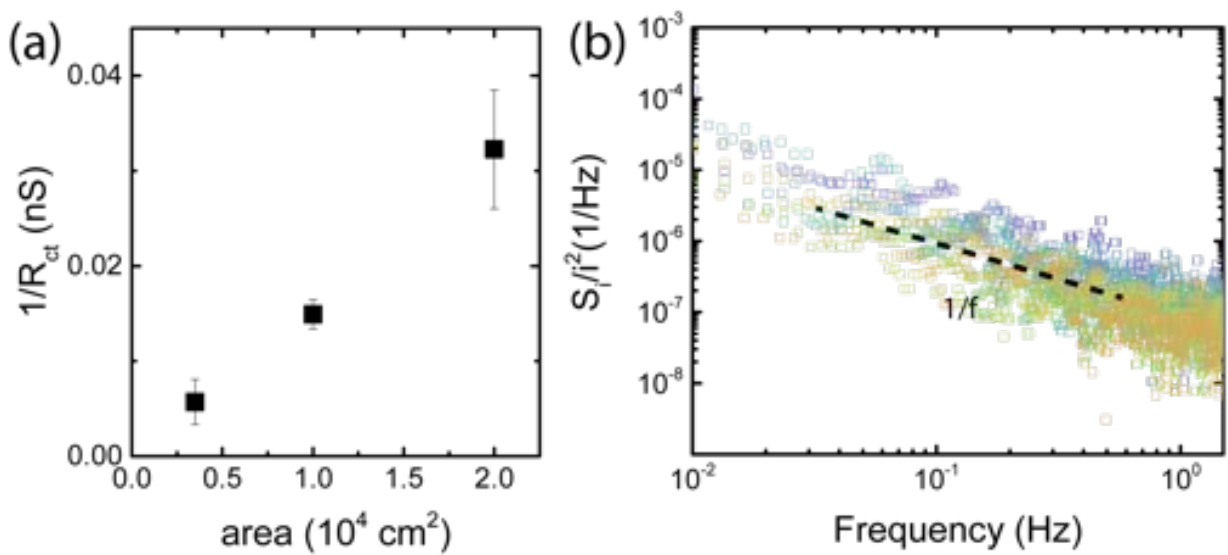

**Figure 3**. (a) The charge-transfer resistance for samples with different area. The error bars were determined by the fitting using the formula based on the theory for electrical double layer, with $R_{ct}$ as a fitting parameter. (b) Normalized charge power spectral density showing the $1/f^{\,1.1\pm0.2}$ dependence. The color of each data set corresponds to the same ionic concentrations as the individual data points in Fig. 2 (a). The dashed line showing the $1/f$ dependence is a guide for the eye.

The best fit value of $R_{ct}$ can be understood by considering electron-transfer processes through graphene defects[29, 35, 36]. By adopting the parameters inferred previously[5] for gold electrodes measured under conditions similar to those used for our graphene electrodes, the standard electron-transfer rate constant of the graphene/solution interface, $k_0$, can be estimated from $R_{ct}$ to be ~ $3.4 \times 10^{-5}$ cm s$^{-1}$. This value agrees with earlier reports for $k_0$ of graphene derived from scanning electrochemical microscopy data[37]. Since $k_0$ values for the pristine basal plane and defect states of graphene are $< 10^{-6}$ cm s$^{-1}$ and $\sim 0.08$ cm s$^{-1}$ respectively[29], the fractional area associated with defects in the graphene region tested here is estimated to be $\sim 0.04\%$. This value agrees well with the defect density inferred from the Raman spectrum[38] presented in Fig. S3 of the Supplemental Material[26]. Further investigation is needed to identify the chemical groups involved in Faradaic charge transfer at the graphene/solution interface. We also quantified the charge-transfer resistance $R_{ct}$ with various electrode-areas. As shown in Fig. 3(a), $R_{ct}$ is inversely proportional to the area of the graphene electrode, with an average specific charge transfer resistance of $6.5 \pm 0.3$ M$\Omega$ cm$^2$.

The characteristic response of the low-level Faradaic current to variations in the electrostatic potential in the solution above the graphene indicates that this methodology is well suited for small-signal measurement of biosensors and biodevices in liquid environments. This methodology presents three key advantages. First, the measurement does not require application of a bias or gate voltage, making it a low power ($\sim$ fW) approach with the potential to be developed into portable or implantable devices. Second, the electrometer-based method has ultimate charge resolution on the femtocoulomb scale, making it significantly more sensitive than conventional approaches based on GFETs. For a GFET fabricated on 300 nm SiO$_2$ substrate with $100 \times 100$ μm$^2$ graphene channel, 1 fC ($\sim$6000 electron charges) corresponds to a surface charge density of $6.25 \times 10^7$ cm$^{-2}$, or $\sim 0.8$ mV shift in the Dirac point voltage, which is very difficult to resolve[11]. Third, the noise level of the electrometer measurement (0.75 fA peak-to-peak) is extremely low compared to other conventional methods. This should enable measurements of charge noise intrinsic to low-noise biosystems, for example the nervous system[39].

The high accuracy of this measurement configuration enables low-level electrical noise measurement of the graphene/solution interface. For an ideal, non-perturbative system with slow Faradaic charge-transfer[29, 30], the power spectral density (PSD) of the current fluctuations, $S_i$, is predicted to show a $1/f^{\alpha}$ dependence, with $\alpha=1$ as a result of the $t^{-1/2}$ time dependency of the ion velocity at the interface[30]. Since the feedback capacitor in the electrometer and the EDL capacitor both contribute negligible charge noise, the measured noise PSD is intrinsic to the solution-graphene system. (See details of extracting $S_i$ in Supplemental Material S4[26]) The normalized current noise PSD $S_i/i^2$ for all ionic concentrations tested collapse into a $1/f^{\alpha}$ characteristic with $\alpha=1.1 \pm 0.2$ as shown in Fig. 3(b), in good agreement with theory for non-perturbative measurement of the electrode-liquid interface[30, 40]. The prefactor for the functional fit to the data in Fig. 3, $\sim 10^{-7}\,\text{Hz}^{-1}$, is within the range predicted by theory (See Supplemental Material S5[26]), $10^{-5} - 10^{-9}$, where the large range is mainly due to uncertainty in estimating the Nernst layer thickness for the system[30, 41].

Our methodology for quantifying $R_{ct}$ and $\sigma_d$ is expected to be appropriate for other electrolytes such as ionic liquids or hydrogels. In those cases, we would expect that the surface charge density and the charge transfer resistance might vary with the identify of the electrolyte, because both the charge of the relevant ionizable groups and the charge-transfer rate, which depends on the interaction between ionizable groups and the active chemicals in the electrolyte, may change.

In summary, we developed an electrometer-based measurement to monitor charge accumulation at the graphene-solution interface, and charge transfer across that interface over time, corresponding to current flows ~ 0.1 – 1.0 pA. The measurement is non-perturbative and does not require the application of a voltage across the graphene-solution interface. Using this method, we determined that the polarity of the adsorbed ions is negative for graphene in aqueous solution system. We quantified the Faradaic current across the graphene-solution interface as a function of ionic strength and found excellent agreement with predictions of the theory of the EDL. The inferred values of the charge transfer resistance and the surface charge density are in good agreement with expectations. The charge noise in the system showed a 1/f dependence, as expected theoretically for the unperturbed graphene-solution interface with slow Faradaic charge transfer process which has not been reported previously. The methodology presented here can be applied to low-level measurement in biodevices and generalized to low-level measurement of other nanomaterials beyond graphene. It should be suitable for the development of low-power, high sensitivity nano-bio sensor devices for use in liquids, such as wearable or implantable systems.

Methods

**Graphene growth.** Copper foil (99.8% purity) is loaded into a four-inch quartz tube furnace and annealed for 30 minutes at 1050 °C in ultra-high-purity (99.999%) hydrogen atmosphere (flow rate 80 sccm; pressure of 850mT at the tube outlet) for removal of oxide residues. Graphene is then deposited by low-pressure chemical vapor deposition (CVD) using methane as a precursor (flow rate 45 sccm, growth time of 60 min).

**Graphene device fabrication.** The graphene-copper growth substrate is coated with a 500 nm layer of poly(methyl methacrylate) (PMMA, MICROCHEM), and the PMMA-graphene film is floated off the surface by immersion in a 0.1M NaOH solution with the graphene-copper growth substrate connected to the cathode of a power supply. The PMMA-graphene film is transferred onto a silicon wafer with an array of 5 nm/40 nm Cr/Au contact electrodes that was previously fabricated using photolithography. After removal of PMMA with acetone, the graphene film is cleaned by annealing at 250 °C in 1000 sccm argon and 400 sccm hydrogen for 1 hour. Then 100 µm x 100 µm graphene electrodes are defined by photolithography (photoresist AZ 5214 E, MICROCHEM) and oxygen plasma etching. Another layer of photoresist SU-8 (MICROCHEM) is then applied to the device, and the passivation layer covering the electrodes is defined via photolithography.

**Corresponding Author**

A. T. Charlie Johnson (cjohnson@physics.upenn.edu)

**Funding Sources**

This work was supported by the Defense Advanced Research Projects Agency (DARPA) and the U. S. Army Research Office under grant number W911NF1010093.

processes studeid by appliction of the linear potential sweep method, Electrochimica Acta, 51 (2006) 2971-2976.